\documentclass[aps,prl,twocolumn,groupedaddress]{revtex4}
\usepackage{graphicx}
\newcommand{\drdez}{$d\rho_{xx}/dE_Z$}
\begin{document}

\title{Spin Transition in the Half-Filled Landau Level}

\author{L. A. Tracy$^1$, J.~P. Eisenstein$^1$, L.~N. Pfeiffer$^2$, and K.~W. West$^2$}

\affiliation{$^1$California Institute of Technology, Pasadena CA 91125 
\\
$^2$Bell Laboratories, Lucent Technologies, Murray Hill, NJ 07974\\}

\date{\today}

\begin{abstract}The transition from partial to complete spin polarization of two-dimensional electrons at half filling of the lowest Landau level has been studied using resistively-detected nuclear magnetic resonance (RDNMR). The nuclear spin-lattice relaxation time is observed to be density independent in the partially polarized phase but to increase sharply at the transition to full polarization. At low temperatures the RDNMR signal exhibits a strong maximum near the critical density.

\end{abstract}


\maketitle
In an ordinary itinerant ferromagnet like iron, electron-electron interactions are responsible for the transition from the paramagnetic to ferromagnetic state.  Similarly, a two-dimensional electron system (2DES) at zero magnetic field is expected to ferromagnetically order at sufficiently low density $n_s$ where interactions dominate over kinetic effects.  At high perpendicular magnetic field $B$ the kinetic energy spectrum of a 2DES is resolved into discrete Landau levels and the importance of electron-electron interactions is vastly enhanced.  Interactions then lead to both ferromagnetic and spin unpolarized phases, depending upon the Landau level filling factor $\nu = hn_s/eB$ and the spin Zeeman energy $E_Z = g\mu_BB$.  For example, in the quantized Hall effect (QHE) states at $\nu = 1$ and 1/3 in the lowest Landau level, the ground state of the 2DES is ferromagnetically ordered, even in the limit of zero Zeeman energy. Remarkably, in the same limit the ground states at $\nu = 2/3$ and 2/5 are unpolarized spin singlets\cite{chakraborty}.  

The situation is less clear at filling factors $\nu$ where no QHE exists. Most important among these is the half-filled Landau level, $\nu = 1/2$.  Numerical exact diagonalization calculations, so effective at incompressible filling factors like $\nu = 1/3$, cannot yet definitively determine the spin polarization of the compressible $\nu =1/2$ system\cite{rezayi}. Composite fermion (CF) theory maps the real electron system at $\nu = 1/2$ onto a system of CFs (electrons with two fictitious flux quanta attached)\cite{jain1}. At the mean field level the CFs experience zero effective magnetic field and display a Fermi surface\cite{HLR,shankar}. In this approximation, the system is a Pauli paramagnet whose spin polarization is set by the ratio $E_Z/E_F$ of the Zeeman energy to the CF Fermi energy. Only if $E_Z > E_F$ is the system fully polarized.  This simple scenario has received qualitative support from several experiments which have demonstrated that the spin polarization at $\nu = 1/2$ is incomplete at low magnetic field and that some kind of transition to complete polarization occurs at higher field\cite{kukushkin,barrett,melinte,freytag,stern,spielman,pinczuk}.  

In this paper we report the results of a careful study of the transition to complete spin polarization at $\nu = 1/2$ as a function of density $n_s$ in a single layer 2DES. The method of resistively-detected nuclear magnetic resonance (RDNMR)\cite{desrat} is used to measure both the nuclear spin--lattice relaxation time $T_1$ and the derivative $d\rho_{xx}/dE_Z$ of the 2DES resistivity with respect to Zeeman energy.  Significant deviations from the simplest CF theory are evident in both of these observables.

The sample used here is a GaAs/AlGaAs heterojunction, modulation-doped with Si. A 2DES, with an as-grown density of $n_s \approx 1.3 \times 10^{11}$ cm$^{-2}$ and low-temperature mobility $\mu \approx 5 \times 10^6$ cm$^2$/Vs lies 600 nm below the sample surface.  The 2DES is laterally patterned into a 500 $\mu m$ wide Hall-bar geometry and is covered by an Al metal gate.  Using the gate $n_s$ may be reduced to below $0.3 \times 10^{11}$ cm$^{-2}$ where the mobility is about $1 \times 10^6$ cm$^2$/Vs.  An 8-turn NMR coil is wound around the sample for applying radio-frequency (RF) magnetic fields $H_1$ parallel to the long axis of the Hall bar.  Large static magnetic fields are applied perpendicular to the 2DES plane.  The RF fields are estimated to be in the $H_1 \sim 0.1~\mu$T range, much less than the typical nuclear dipolar fields $H_d \sim 100~\mu$T.  The electron temperatures quoted here are corrected for heating due to the RF field.  The small, but readily measurable, temperature dependence of the 2DES resistivity at $\nu = 1/2$ allows for an accurate calibration of the heating effect.

\begin{figure}[t]
\centering
\includegraphics[width=3.5 in,bb= 73 73 417 338]{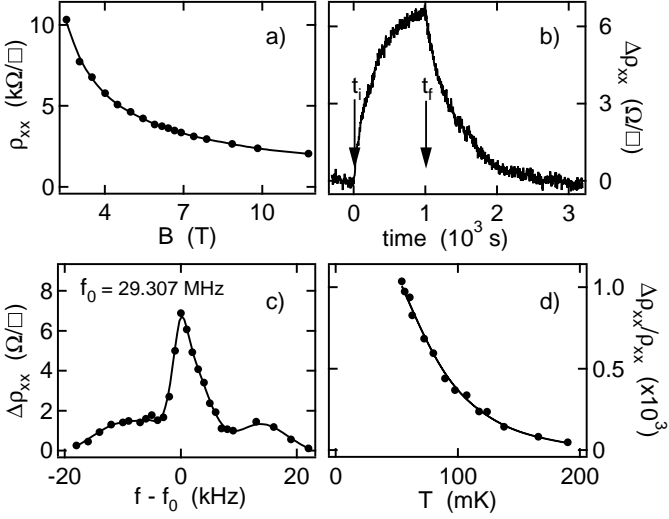}
\caption{a) Resistivity at $\nu = 1/2$ versus magnetic field at $T$ = 45 mK. b) Typical response of resistivity at $\nu = 1/2$ when RF frequency is brought on and off resonance with the $^{75}$As NMR line. c) Typical RDNMR lineshape. d) Temperature dependence of RDNMR signal at peak of line at $\nu =1/2$ and $B = 4.01$ T.}
\label{fig1}
\end{figure}
The RDNMR technique relies on the hyperfine interaction.  A finite nuclear spin polarization $\xi_N$ creates an effective magnetic field $B_N$ which modifies the electronic Zeeman energy:  $E_Z = g \mu_B (B+B_N)$.  For electrons in GaAs, where $g = -0.44$, $B_N$ = -5.3 T if all three spin-3/2 nuclear species ($^{69}$Ga, $^{71}$Ga, and $^{75}$As) are fully polarized\cite{paget}.  Reduction of the nuclear polarization, via NMR excitation, increases $E_Z$ and will alter the 2DES resistivity if $d\rho_{xx}/dE_Z \ne 0$. 

Low temperature magneto-transport measurements on this sample reveal the integer and fractional QHE states ($e.g.$ at $\nu =1/3$, 2/5, etc.) typically found in other 2DES samples of comparable mobility.  Figure 1a shows that the longitudinal resistivity $\rho_{xx}$ at $\nu = 1/2$ increases steadily as $n_s$  (and thus the magnetic field where $\nu =1/2$) is reduced.  This is expected since the random disorder potential remains fixed.  At the lowest density $\rho_{xx}\sim 0.4 h/e^2$.  The data in Fig. 1a offer no hint of any transition involving the spin degree of freedom.

Figure 1b shows a typical response of the resistivity at $\nu =1/2$ to the application of an RF magnetic field tuned to the $^{75}$As NMR line. Prior to time $t_i$ the RF field is on, but set to a frequency 47 kHz below the NMR line, here at 32.747 MHz.  The nuclear spin polarization is at its thermal equilibrium value.  At time $t_i$ the RF frequency is brought onto resonance and the nuclear spin polarization begins to fall. Simultaneously, the resistivity $\rho_{xx}$ begins to increase and approach a saturation value.  Both the rate of increase and the saturation value are RF power dependent.  At time $t_f$ the RF frequency is returned to its off-resonance value and $\rho_{xx}$ decays back to its equilibrium value.  Both the rise and the fall of the RDNMR signal are exponential in time. According to the Bloch equations\cite{abragam}, the rise and fall times are $\tau_{rise} = T_1/(1+\omega_R^2T_1T_2)$ and $\tau_{fall} = T_1$, with $\omega_R \propto H_1$ the Rabi frequency, $T_1$ the nuclear spin--lattice relaxation time, and $T_2$ the nuclear spin dephasing time. The rise and fall times also determine the fractional change of the nuclear spin polarization of $^{75}$As induced by the RF: $\Delta \xi_N/\xi_N = \tau_{rise}/\tau_{fall}-1$.  The fractional change in nuclear polarization was kept roughly constant at $\Delta \xi_N/\xi_N \approx -0.47$. 

Figure 1c shows the amplitude of the RDNMR signal, $\Delta \rho_{xx}$, versus frequency for the $^{75}$As NMR line.  The main peak is asymmetric and roughly 5 kHz wide.  The satellite peaks are due to quadrupole splitting.  The width and asymmetry of the main peak are most likely due to variation of the Knight shift associated with the shape of the electronic wavefunction perpendicular to the 2DES plane. Nuclei near the peak in the wavefunction experience a larger Knight shift and a shorter $T_1$ time than those in the tails.  In the following, all RDNMR data shown were taken at the peak of the $^{75}$As NMR line.

Finally, Fig. 1d shows a typical temperature dependence of the RDNMR signal at $\nu = 1/2$. These data, taken at $B = 4.0$ T, reflect both the equilibrium nuclear polarization, $\xi_N \sim B/T$, and the temperature dependence of the derivative $d\rho_{xx}/dE_Z$ of the 2DES.

\begin{figure}[t]
\centering
\includegraphics[width=3.0 in,bb= 73 73 314 355]{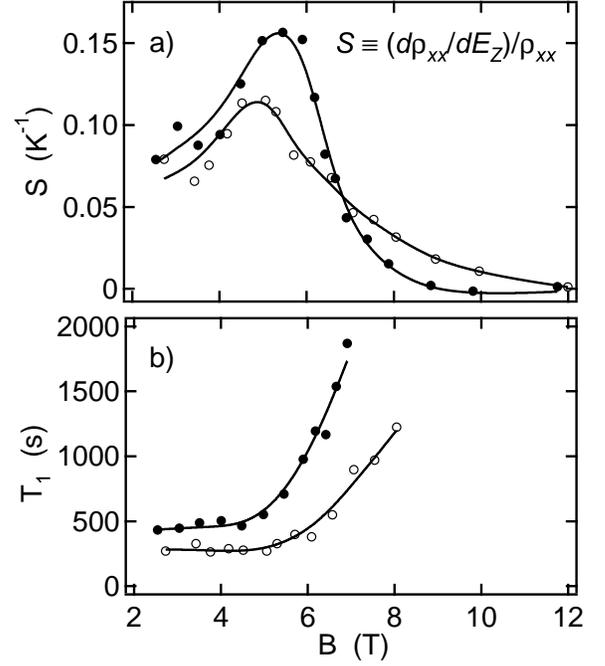}
\caption{a) Derivative $S \equiv (d\rho_{xx}/dE_Z)/\rho_{xx}$ vs. magnetic field at $\nu = 1/2$ at $T = 45$ mK (solid dots) and  100 mK (open dots). b) Nuclear spin-lattice relaxation time $T_1$ vs. field at $\nu = 1/2$ at the same temperatures.}
\label{fig2}
\end{figure}
Figure 2 contains our most important results.  Figure 2a displays $(d\rho_{xx}/dE_Z)/\rho_{xx}$ versus magnetic field (and thus density) at $\nu = 1/2$ at two temperatures: $T =$ 45 mK and 100 mK.  These data were obtained by dividing the observed NMR-induced changes in resistivity $\Delta \rho_{xx}$ by the concomitant change $\Delta E_Z$ in the Zeeman energy.  $\Delta E_Z$ was computed from the fractional change in the nuclear polarization $\Delta \xi_N /\xi_N$ deduced from the measured $\tau_{rise}$ and $\tau_{fall}$ of the RDNMR signal and the known\cite{paget} hyperfine parameters and equilibrium polarization $\xi_N$ of $^{75}$As.  The $T = 45$ mK data show a dramatic transition from a low field (and density) regime where $d\rho_{xx}/dE_Z$ is substantial to a high field regime where it essentially vanishes.  A strong peak is apparent in $d\rho_{xx}/dE_Z$ in the transition region.  The $T = 100$ mK data show a similar transition, albeit substantially smeared out.

Figure 2b displays the field/density dependence of the $T_1$ time at $\nu = 1/2$.  A clear transition is again observed.  At low fields and densities $T_1$ is relatively short and density independent.  Above a critical field, however, $T_1$ begins to rise rapidly. At $T =$ 45 mK this rise is more pronounced than at 100 mK. Figure 3 shows the temperature dependence of $T^{-1}_1$ at $B \approx 3.0$, 4.0, 5.0, and 6.4 T, corresponding to densities of $n_s = 0.36$, 0.48, 0.60, and 0.78 $\times 10^{11}$ cm$^{-2}$.  The figure shows that nuclear spin--lattice relaxation rate $T^{-1}_1$ at $\nu = 1/2$ is linear in temperature, $T^{-1}_1 = \alpha T + \beta$, and independent of magnetic field and density up to about $B = 5$ T.  At higher fields and densities the temperature dependence of $T^{-1}_1$ becomes non-linear.  At $B = 6.4$ T the dependence is essentially quadratic, albeit with an offset in the zero temperature limit comparable to the offset $\beta$ seen at lower density. 

\begin{figure}[t]
\centering
\includegraphics[width=3.25 in,bb= 0 0 286 194]{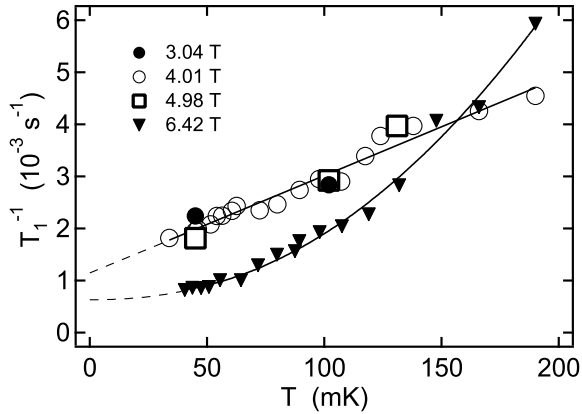}
\caption{Temperature dependence of $T^{-1}_1$ at $\nu = 1/2$. The solid lines are straight line and parabolic fits to the data, with the dashed portions being extrapolations to $T = 0$.}
\label{fig2}
\end{figure}
The data in Fig. 2a strongly suggest that the 2DES at $\nu = 1/2$ undergoes a transition from partial spin polarization at low density and magnetic field to complete polarization at high density and field.  If the 2DES is fully spin polarized, increasing $E_Z$, via NMR, cannot further increase the polarization.  No effect on the resistivity $\rho_{xx}$ would be expected.  The collapse of the RDNMR signal at high magnetic field is consistent with this.  Conversely, if the 2DES spin polarization is incomplete, an NMR-induced increase of the Zeeman energy will increase the electronic polarization and a change in the resistivity can be expected.  Our data demonstrate that $d\rho_{xx}/dE_Z > 0$ in the low field, partially polarized regime\cite{nonlin}.  

Figures 2b and 3 provide strong independent support for a spin transition. At low fields and densities $T_1$ is relatively short and possesses a Korringa-like temperature dependence. Since the nuclear Larmor energy is negligible ($\sim 40$ MHz $\approx$ 2 mK), hyperfine-assisted nuclear spin flops can only occur if essentially zero energy electron spin flips are possible.  As in ordinary metals, these facts imply that the Fermi level of the 2DES lies in both spin branches simultaneously; i.e. the 2DES is at most partially spin polarized, even at $T = 0$\cite{phasetransition}.  Above about $B = 5$ T however, $T_1$ rises rapidly and assumes a non-Korringa temperature dependence.  The Zeeman energy now exceeds the Fermi energy.  At $T = 0$ the 2DES would be fully spin polarized and exhibit an extremely long $T_1$, likely dominated by non-electronic processes. Consistent with our observations, at non-zero temperature both partial spin polarization and a finite $T_1$ will persist with increasing density until $E_Z- E_F >> k_BT$.

A qualitative model of the spin polarization transition at $\nu = 1/2$ follows from CF theory\cite{jain2,shankar}.  In mean field approximation there will be two Fermi surfaces, one for each CF spin species, provided that $E_Z$ is less than the majority spin Fermi energy.  Assuming the bands are parabolic and defined by a single effective mass parameter $m^*_{CF}$, the spin polarization $\xi_{CF} = (n_{\uparrow}-n_{\downarrow})/n_s$ is just $\xi_{CF}=|g|m^*_{CF}/m_0$, where $m_0$ is the mass of a bare electron.  This formula follows from the fact that both $E_Z$ and the total electron density at $\nu = 1/2$, $n_s = eB/2h$, are proportional to the magnetic field $B$. Thus, if $m^*_{CF}/m_0 < 1/|g| \approx 2.3$ the 2DES will be partially spin polarized.  Since $m^*_{CF}$ is presumed to scale with field as $m^*_{CF} \propto B^{1/2}$, partial spin polarization persists only up to a critical field (or density) beyond which the 2DES is fully polarized.  Figure 2 suggests that the mid-point of this transition is around $B = 6$ T, suggesting $m^*_{CF}/m_0 \approx 0.9 B^{1/2}$, somewhat larger than the theoretical value $m^*_{CF}/m_0 \approx 0.6 B^{1/2}$ given by Park and Jain\cite{jain2}. 

Despite the ready ability of this simplest CF theory to explain the existence of a spin transition at $\nu = 1/2$, it fails to fully account for our observations.  For example, in this theory the nuclear spin--lattice relaxation rate $T^{-1}_1$ in the partially polarized phase should follow a 2D version of the Korringa law commonplace in ordinary metals\cite{shankar2}.  Indeed, the linear temperature dependence of $T^{-1}_1$ shown in Fig. 2a is consistent with this (the offset $\beta$ arises, in part, from finite temperature corrections to the 2D Korringa law\cite{shankar2} but almost certainly also from nuclear spin diffusion\cite{diffusion}). In ordinary metals $T_1$ is inversely proportional to the square of the density of states at the Fermi level.  In the present $\nu = 1/2$ CF case this implies $T_1 \propto (m^*_{CF})^{-2} \propto B^{-1}$.  Figs. 2a and 3 clearly show that this dependence is not observed, with $T_1$ being independent of magnetic field in the partially polarized phase.  Of course, the Korringa rate also depends upon the shape of the 2DES subband wavefunction and this becomes `thinner and taller' as the density is increased. As the wavefunction thins, the Coulomb interaction between electrons is modified.  Interestingly however, since the filling factor is fixed at $\nu = 1/2$, the decreasing magnetic length $\ell$ overcompensates for the thinning effect and the net result is a relative \emph{softening} of the repulsion between electrons\cite{zhang}.  This enhances $m^*_{CF}$ beyond $B^{1/2}$ and forces the theoretical $T_1$ to fall faster than $B^{-1}$.  In addition, as the wavefunction becomes taller with increasing density, the hyperfine matrix elements increase and further suppress $T_1$. Hence, these finite thickness effects\cite{LDA} only worsen the comparison between the observed $T_1$ and CF theory. 

The simple model of the CF spin polarization presented above also does not explain the strong peak in \drdez\ near the critical density. However, the similarity between CF's at $\nu = 1/2$ and ordinary 2D electrons at zero (perpendicular) magnetic field may offer clues to its origin. In that case spin polarization can be induced without Landau quantization by applying a strong magnetic field $B_{||}$ parallel to the 2D plane.  Das Sarma and Hwang\cite{dassarma} have calculated the magneto-resistance $\rho_{xx}(B_{||})$ under the assumption that the resistivity is dominated by screened impurity scattering.  In their theory $\rho_{xx}(B_{||})$ rises by a density-dependent factor as $B_{||}$ increases and the 2DES spin polarizes. Once fully polarized $\rho_{xx}$ becomes independent of $B_{||}$ in thin 2D systems. Interestingly, the derivative $d\rho_{xx}/dB_{||}$, which is proportional to \drdez\ in thin 2D systems, exhibits a peak near the critical field at low temperatures.  Numerous experiments ($e.g.$ on 2D electrons Si\cite{vitkalov} and 2D holes in GaAs\cite{tutuc,gao}) have observed behavior qualitatively like this, although there is wide variation in the magnitude of the resistance change and the peak in $d\rho_{xx}/dB_{||}$ is generally quite broad.  Although this suggests that the peak in \drdez\ we observe at $\nu = 1/2$ might result from physics similar to that which governs the in-plane magneto-resistance of ordinary 2D electrons, the analogy remains speculative since there is as yet no transport theory of partially polarized CF's.  Indeed, even the $sign$ of the RDNMR signal that we observe at $\nu =1/2$ has yet to be understood.   

An alternative, if still more speculative, possibility is that the peak in \drdez\ is an indication that the transition to full spin polarization at $\nu = 1/2$ is weakly first-order and that near the critical point the 2DES phase separates into domains of partial and complete spin polarization. Static fluctuations in the 2DES density (due to the random donor distribution) would encourage such phase separation and, in analogy to exchange-driven ferromagnetism, residual interactions between the CFs might force the spin polarization to jump discontinuously\cite{jain3} from $\xi_{CF}<1$ to $\xi_{CF}=1$ at the domain walls. Our results imply that an NMR-induced increase of the Zeeman energy would both increase the resistivity of the partially polarized regions \emph{and} move the domain walls so as to reduce the size of such regions.  While the former effect can produce a RDNMR signal even in a homogeneous partially polarized 2DES, domain wall rearrangement will create an independent contribution to the RDNMR response only in a phase-separated system near the critical point.  This contribution could explain the peak in $d\rho_{xx}/dE_Z$.

In conclusion, we have used RDNMR to study the spin polarization transition in a 2DES at $\nu =1/2$.  A simple model of non-interacting composite fermions fails to explain the behavior of the nuclear $T_1$ and the derivative $d\rho_{xx}/dE_Z$ of the 2DES resistivity with respect to Zeeman energy near the critical point.   

We acknowledge helpful conversations with N. Bonesteel, S. Das Sarma, G. Murthy, E. Rezayi, R. Shankar, S. Simon, and A. Stern.  This work was supported by the DOE under Grant No. DE-FG03-99ER45766 and the NSF under Grant No. DMR-0552270.     


\begin{references}
\bibitem{nonlin} This result holds so long as the current used to measure $\rho_{xx}$ is not too large.  At high current the RDNMR signal becomes non-linear and ultimately changes sign (See Ref. \cite{stern}).  In this regime the nuclei are being pumped out of equilibrium by the current. This in turn suggests that the electronic polarization may not be in equilibrium.
\bibitem{LDA} We have verified both of these effects using local density approximation calculations of the subband wavefunction shape and its effect on the Coulomb pseudopotentials.
\end{references}
\end{document}